\documentclass[conference,10pt]{IEEEtran}
\IEEEoverridecommandlockouts
\usepackage[utf8]{inputenc}
\usepackage[T1]{fontenc}
\usepackage{amsmath,amssymb,amsfonts}
\usepackage{graphicx}
\usepackage{textcomp}
\usepackage{pgfplots}
\usepackage{scalefnt}

\newcommand{\diag}{\mathop{\rm diag}}
\newcommand{\tr}{\mathop{\rm tr}}
\newcommand{\tx}{\mathop{\rm Tr}}
\newcommand{\nondiag}{\mathop{\rm nondiag}}
\graphicspath{{figures/}}
\def\BibTeX{{\rm B\kern-.05em{\sc i\kern-.025em b}\kern-.08em
    T\kern-.1667em\lower.7ex\hbox{E}\kern-.125emX}}
\begin{document}

\title{Study of Joint Automatic Gain Control and MMSE Receiver Design Techniques for Quantized Multiuser
Multiple-Antenna Systems\\
\thanks{The authors would like to thank the CAPES
Brazilian agency for funding.}
}

\author{\IEEEauthorblockN{T. E. B. Cunha$^{1}$, R. C. de Lamare$^{1}$, T. N. Ferreira$^{2}$, T. Hälsig$^{3}$}
\IEEEauthorblockA{$^{1}$Centre for Telecommunications Studies (CETUC), Pontifical Catholic University of Rio de Janeiro, Brazil \\
$^{2}$Engineering School, Fluminense Federal University (UFF), Niterói, RJ, Brazil \\
$^{3}$Institute for Communications Engineering, Universität der Bundeswehr München, Germany \\
Email: \{thiagoelias, delamare\}@cetuc.puc-rio.br, tadeu\_ferreira@id.uff.br, tim.haelsig@unibw.de }
}

\maketitle
\begin{abstract}

This paper presents the development of a joint optimization of an
automatic gain control (AGC) algorithm and a linear \textit{minimum
mean square error} (MMSE) receiver for multi-user multiple input
multiple output (MU-MIMO) systems with coarsely quantized signals.
The optimization of the AGC is based on the minimization of the
\textit{mean square error} (MSE) and the proposed receive filter
takes into account the presence of the AGC and the effects due to
quantization. Moreover, we provide a lower bound on the capacity of
the MU-MIMO system by deriving an expression for the achievable
rate. The performance of the proposed Low-Resolution Aware MMSE
(LRA-MMSE) receiver and AGC algorithm is evaluated by simulations,
and compared with the conventional MMSE receive filter and
Zero-Forcing (ZF) receiver using quantization resolution of 2, 3, 4
and 5 bits.
\end{abstract}

\begin{IEEEkeywords}
Coarse Quantization, AGC, MU-MIMO detection, MMSE receiver
\end{IEEEkeywords}

\section{INTRODUCTION}
In 5G celular systems, high data rates, reliable links, low cost and
power consumption are key requirements. Multiple-input
multiple-output (MIMO) systems in wireless communications provide
significant improvements in wireless link reliability and achievable
rates. However, as the number of antennas scales up, the energy
consumption and circuit complexity increases accordingly
\cite{b1,b3,Landau2017,ZShao,Landau2018}. For example, the energy
consumption of an analog to digital converter (ADC) grows
exponentially as a function of the quantization resolution. To
reduce circuit complexity and save energy, novel transmission
approaches employ low resolution ADCs, which generate significant
nonlinear distortion and, thus, require new signal processing
techniques to provide reliable data transmission. Several studies
present detection methods that deal with signals quantized with few
bits of resolution \cite{b1,b2,b3,ZShao}.

Automatic gain control (AGC) is the process of adjusting the analog
signal level to the dynamic range of the analog to digital converter
(ADC) in order to minimize the signal distortion due to the
quantization \cite{agc_effects}. The use of an AGC is important in
applications where the received power varies over time, as is the
case in mobile scenarios. Proper AGC design becomes especially
important for low resolution ADCs. Although there are many articles
on quantization in MIMO systems in the literature, few address the
design of AGCs. In \cite{b1}, the authors presented a modified MMSE
receiver that takes into account the quantization effects in a MIMO
system but they do not take into account the presence of an AGC. The
effects of an AGC on a quantized MIMO system with a standard
Zero-Forcing filter at the receiver were examined in \cite{b2}.
However, the authors have not optimized the AGC algorithm nor used a
detector that considers the quantization effects.

This work presents a framework for jointly designing the AGC and a
linear receive filter according to the MMSE criterion for a
large-scale MU-MIMO system operating with coarsely quantized
signals. The procedure consists of computing the modified MMSE
receiver presented in  \cite{b1} and, after that, computing the
derivative of the cost function that takes into account the presence
of the AGC in order to obtain the optimal AGC coefficients. Then, a
Low-Resolution Aware MMSE (LRA-MMSE) receiver that considers both
quantization effects and the AGC is derived. A lower bound on the
capacity of this system is investigated and an expression to compute
the achievable rates is developed.

\textit{Notation}: Vectors and matrices are denoted by lower and
upper case italic bold letters. The operators $(\cdot)^T$,
$(\cdot)^H$ and $\tr(\cdot)$ stand for transpose, Hermitian
transpose and trace of a matrix, respectively. $\mathbf{1}$ denotes
a column vector of ones and $\mathbf{I}$ denotes an identity matrix.
The operator $E[\cdot]$ stands for expectation with respect to the
random variables and the operator $\odot$ corresponds to the
Hadamard product. Finally, $\diag(\mathbf{A})$ denotes a diagonal
matrix containing only the diagonal elements of $\mathbf{A}$ and
$\nondiag(\mathbf{A})=\mathbf{A}-\diag(\mathbf{A})$.

\section{System Description}

A large-scale uplink MU-MIMO system \cite{mmimo,wence} consisting of
a base station (BS) with $N_R$ receive antennas, and $K$ users
equipped with $N_T$ transmit antennas each is considered. At each
time instant $i$, each user transmits $N_T$ symbols which are
organized into a $N_T \times 1$ vector
$\mathbf{x}_k[i]=[x_1[i],x_2[i],...,x_{N_T}[i]]^T$. Each entry of
the vector $\mathbf{x}_k[i]$ is a symbol taken from the modulation
alphabet $\mathbb{A}$. The symbol vector is then transmitted through
flat fading channels and corrupted by additive white Gaussian noise
(AWGN). The received signal collected by the receive antennas at the
BS is given by the following equation:
    \begin{eqnarray}
    \mathbf{y}[i]=\sum_{k=1}^{K} \mathbf{H}_{k}\mathbf{x}_{k}[i]+\mathbf{n}[i]= \mathbf{H}\mathbf{x}[i]+\mathbf{n}[i],
    \label{eq:2}
    \end{eqnarray}
\noindent where $\mathbf{y}[i] \in \mathbb{C}^{N_R \times 1}$, and $\mathbf{H}_k \in \mathbb{C}^{N_R \times N_T}$ is a matrix that contains the complex channel gains from the $N_T$ transmit antennas of user $k$ to the $N_R$ receive antennas of the BS. The $N_R \times 1$ vector $\mathbf{n}[i]$ is a zero-mean complex circular symmetric Gaussian noise vector with covariance matrix $E[\mathbf{n}[i]\mathbf{n}^H[i]]=\sigma_n^2\mathbf{I}$. $\mathbf{H}$ is a $N_R \times KN_T$ matrix that contains the coefficients of the flat fading channels between the transmit antennas of the $K$ users and the receive antennas of the BS. The symbol vector $\mathbf{x}[i]=[\mathbf{x}_1[i],...,\mathbf{x}_k[i],...,\mathbf{x}_{K}[i]]^T$ contains all symbols that are transmitted by the users at time instant $i$. The channel state information (CSI) is assumed to be unknown to the users at the transmit side. Therefore, we assume the same symbol energy per user and transmit antenna, i.e. $E[\mathbf{x}\mathbf{x}^H]=\sigma_ x^2\mathbf{I}$.

\begin{figure}[!h]
\centering
\includegraphics[scale=0.45]{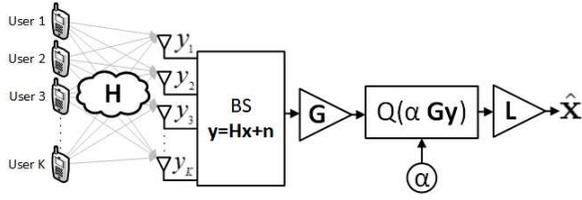}
\caption{An uplink quantized MU-MIMO system}
\label{fig:system_model}
\end{figure}

As depicted in Fig.\ref{fig:system_model}, before to be quantized,
$\mathbf{y}$ is jointly pre-multiplied by the clipping level factor
$\alpha$ and the AGC matrix $\mathbf{G}$ to minimizes the granular
and the overload distortions. After that, the product
$\alpha\mathbf{G}\mathbf{y}$ is quantized and the estimation of the
transmitted symbol vector $\mathbf{x}$ is performed by the LRA-MMSE
receiver, represented by $\mathbf{L}$. The estimated symbol vector
is represented by $\hat{\mathbf{x}}$. The real, $y_{i,R}$, and
imaginary, $y_{i,I}$, parts of the complex received signal at each
antenna are quantized separately by uniform $b$-bit resolution ADCs.
Therefore, the resulting quantized signals read as
\begin{eqnarray}
r_{i,l}=Q(y_{i,l})= y_{i,l}+q_{i,l},\hspace{4pt} l \in \{R,I\}, \hspace{4pt} 1 \leq i \leq N_R,
\end{eqnarray}
\noindent where $Q(\cdot)$ denotes the quantization operation and $q_{i,l}$ is the resulting quantization error.

The distortion factor indicates the relative amount of quantization
noise generated by the quantizer, and is given by
$\rho_q^{(i,l)}=\sigma_{q_{i,l}}^2/\sigma_{y_{i,l}}^2$, where
$\sigma_{q_{i,l}}^2$ is the variance of the quantizer error and,
$\sigma_{y_{i,l}}^2$ is the variance of the input $y_{i,l}$. This
factor depends on the number of quantization bits  $b$, the
quantizer type, and the probability density function of $y_{i,l}$
\cite{b1}. In this work the scalar uniform quantizer processes the
real and imaginary parts of the input signal $y_{i,l}$ in a range
$\pm \frac{\sqrt{b}}{2}$. With a high number of antennas the input
signals of the quantizer are approximately Gaussian distributed and
they undergo nearly the same distortion factor $\rho_q$.  It was
shown in \cite{asymptotic} for the uniform quantizer case, that
optimal quantization step $\Delta$ for a Gaussian source decreases
as $\sqrt{b}2^{-b}$ and that $\rho_q$ is asymptotically well
approximated by $\frac{\Delta^2}{12}$.

\section{Proposed Joint AGC and Linear MMSE Receiver Design}

The procedure for joint optimization of the AGC algorithm and the
LRA-MMSE receiver carries out alternating computations between the
AGC and the LRA-MMSE receiver. The first step consists of computing
an LRA-MMSE receive filter that considers the quantization effects.
Other approaches to computing linear MMSE or related filters
\cite{jidf,jio,jiomimo,rrmber} can also be considered. After that,
we compute the derivative of the cost function to obtain the optimal
AGC coefficients. Then, an updated LRA-MMSE receiver is computed.

\subsection{Linear LRA-MMSE Receive Filter Design}

In this first step we do not consider the presence of the AGC in the
system. Thus, the received signal after the quantizer is expressed,
with the \textit{Bussgang} decomposition \cite{bussgang}, as a
linear model $\mathbf{r}=\mathbf{y}+\mathbf{q}$. To develop the
linear receive filter $\mathbf{W}$ that minimizes the MSE we use the
\textit{Wiener-Hopf} equations:
\begin{eqnarray}
\mathbf{W}=\mathbf{R}_{xr}\mathbf{R}_{rr}^{-1},
\label{eq:wiener}
\end{eqnarray}
where the auto-correlation matrix $\mathbf{R}_{rr}$ is given by
\begin{eqnarray}
\mathbf{R}_{rr}=E[\mathbf{r}\mathbf{r}^H]=\mathbf{R}_{yy}+\mathbf{R}_{yq}+\mathbf{R}_{yq}^H+\mathbf{R}_{qq},
\label{eq:Rrr}
\end{eqnarray}
and the cross-correlation matrix $\mathbf{R}_{xr}$ can be expressed as
\begin{eqnarray}
\mathbf{R}_{xr}=E[\mathbf{x}\mathbf{r}^H]=\mathbf{R}_{xy}+\mathbf{R}_{xq}
\label{eq:Rxr}
\end{eqnarray}
We get the auto-correlation matrix $\mathbf{R}_{yy}$ and the cross-correlation matrix $\mathbf{R}_{xy}$ directly from the MIMO model as
\begin{eqnarray}
\mathbf{R}_{yy}=E[\mathbf{y}\mathbf{y}^H]=\mathbf{H}\mathbf{R}_{xx}\mathbf{H}^H+\mathbf{R}_{nn} \label{eq:Ryy}
\end{eqnarray}
and,
\begin{eqnarray}
\mathbf{R}_{xy} = E[\mathbf{x}\mathbf{y}^H] = \mathbf{R}_{xx}\mathbf{H}^H
\label{eq:Rxy}
\end{eqnarray}
To compute (\ref{eq:Rrr}) and (\ref{eq:Rxr}) we need to obtain the covariance matrices $\mathbf{R}_{yq}$, $\mathbf{R}_{qq}$ and $\mathbf{R}_{xq}$ as a function of the channel parameters and the distortion factor $\rho_q$. The procedure of how to obtain these matrices was developed in \cite{b1} and we will use some of these results in this work. The cross-correlation between the received signal vector and the quantization error is approximated by
\begin{eqnarray}
\mathbf{R}_{yq} &\approx& -\rho_q\mathbf{R}_{yy}
\label{eq:Ryq}
\end{eqnarray}

The covariance matrix of the quantization error is deduced from
\begin{eqnarray}
\mathbf{R}_{qq} &\approx& \rho_q \diag(\mathbf{R}_{yy})+\rho_q^2 \nondiag(\mathbf{R}_{yy}) \nonumber \\
&=& \rho_q \mathbf{R}_{yy}-(1-\rho_q)\rho_q \nondiag(\mathbf{R}_{yy}),
\label{eq:Rqq}
\end{eqnarray}

\noindent and the cross-correlation matrix between the desired signal vector and the quantization error can be obtained by
\begin{eqnarray}
\mathbf{R}_{xq}= -\rho_q\mathbf{R}_{xy} \label{eq:Rxq}
\end{eqnarray}

Substituting (\ref{eq:Rxq}) in (\ref{eq:Rxr}) we get
\begin{eqnarray}
\mathbf{R}_{xr}=(1-\rho_q)\mathbf{R}_{xy}
               \label{eq:Rxr_2}
\end{eqnarray}
\noindent and substituting (\ref{eq:Ryy}), (\ref{eq:Ryq}) and (\ref{eq:Rqq}) in (\ref{eq:Rrr}) we get
\begin{eqnarray}
\mathbf{R}_{rr} \approx (1-\rho_q)(\mathbf{R}_{yy}-\rho_q\nondiag(\mathbf{R}_{yy}))
\label{eq:Rrr_2}
\end{eqnarray}

Finally, by substituting (\ref{eq:Rxr_2})  and (\ref{eq:Rrr_2}) in
(\ref{eq:wiener}) we get the expression of the LRA-MMSE receive
filter for a quantized MU-MIMO system. As shown in \cite{b1}, we can
write this solution as
    \begin{eqnarray}
    \mathbf{W}=\mathbf{R}_{xy}(\mathbf{R}_{yy}-\rho_q\nondiag(\mathbf{R}_{yy}))^{-1}
    \end{eqnarray}

With the presence of the AGC the expression of the received vector
changes and can be computed by
$\mathbf{z}=\mathbf{G}\mathbf{y}+\mathbf{q}$. With the same
procedure as before, the MMSE filter with AGC can be computed
through the \textit{Wiener-Hopf} equations as
$\mathbf{L}=\mathbf{R}_{xz}\mathbf{R}_{zz}^{-1}$. The
auto-correlation matrix $\mathbf{R}_{zz}$ and the cross-correlation
matrix $\mathbf{R}_{xz}$ can be computed similarly by
\begin{eqnarray}
\mathbf{R}_{zz}&=&E[\mathbf{z}\mathbf{z}^H]=\mathbf{G}\mathbf{R}_{yy}\mathbf{G}+\mathbf{G}\mathbf{R}_{yq}+\mathbf{R}_{yq}^H\mathbf{G}+\mathbf{R}_{qq}  \nonumber \\
\mathbf{R}_{xz}&=&E[\mathbf{x}\mathbf{z}^H]=\mathbf{R}_{xy}\mathbf{G}+\mathbf{R}_{xq} \nonumber
\label{eq:Rzz}
\end{eqnarray}
We note that nonlinear receiver structures can also be considered
following the approaches reported in
\cite{spa,mfsic,mbdf,did,bfidd}.

\subsection{AGC Design}
    In \cite{b2}, the authors proposed a standard AGC algorithm by using a diagonal matrix $\mathbf{G}$ with real coefficients. This matrix is used to compensate the gain differences of the propagation channel and involves a search over a transmitted symbol alphabet. This approach is very computationally demanding in an environment with a high number of antennas. Writing $\mathbf{G}$ as $\diag(\mathbf{g})$, where $\mathbf{g}$ is a column vector with the diagonal elements of $\mathbf{G}$, the proposed AGC algorithm is based on the minimization of the cost function:
\begin{eqnarray}
\varepsilon &=&E[||\mathbf{x}-\hat{\mathbf{x}}||^2] \nonumber \\
           &=&E[||\mathbf{x}-\mathbf{W}(\alpha\diag(\mathbf{g})\mathbf{y}+\mathbf{q})||^2]
\end{eqnarray}

\noindent and since $\mathbf{G}$ is a diagonal matrix with real coefficients we have $\diag(\mathbf{g})^H=\diag(\mathbf{g})$. Then,
\begin{align}
\varepsilon = &\tr(\mathbf{R}_{xx}-\alpha\mathbf{R}_{xy}\diag(\mathbf{g})\mathbf{W}^H-\mathbf{R}_{xq}\mathbf{W}^H \nonumber \\
           &-\alpha\mathbf{W}\diag(\mathbf{g})\mathbf{R}_{xy}^H+\alpha^2\mathbf{W}\diag(\mathbf{g})\mathbf{R}_{yy}\diag(\mathbf{g})\mathbf{W}^H\nonumber \\
           &+\alpha\mathbf{W}\diag(\mathbf{g})\mathbf{R}_{yq}\mathbf{W}^H-\mathbf{W}\mathbf{R}_{xq}^H \nonumber \\
           &+\alpha\mathbf{W}\mathbf{R}_{yq}^H\diag(\mathbf{g})\mathbf{W}^H+\mathbf{W}\mathbf{R}_{qq}\mathbf{W}^H) \nonumber
\end{align}
    To obtain the optimum $\mathbf{G}$ matrix we compute the derivative of the MSE cost function with respect to $\diag(\mathbf{g})$, equate the derivative terms to zero and solve for $\mathbf{g}$:

    \begin{align}
        \frac{\partial \varepsilon}{\partial \mathbf{g}}=&-\alpha\underbrace{\frac{\partial}{\partial \mathbf{g}}  \tr(\mathbf{R}_{xy}\diag(\mathbf{g})\mathbf{W}^H}_{I})-\alpha\underbrace{\frac{\partial}{\partial \mathbf{g}}  \tr(\mathbf{W}\diag(\mathbf{g})\mathbf{R}_{xy}^H}_{II}) \nonumber \\
           &+\alpha^2\underbrace{\frac{\partial}{\partial \mathbf{g}}  \tr(\mathbf{W}\diag(\mathbf{g})\mathbf{R}_{yy}\diag(\mathbf{g})\mathbf{W}^H}_{III})\nonumber \\
           &+\alpha\underbrace{\frac{\partial}{\partial \mathbf{g}}  \tr(\mathbf{W}\diag(\mathbf{g})\mathbf{R}_{yq}\mathbf{W}^H}_{IV}) \nonumber \\
           &+\alpha\underbrace{\frac{\partial}{\partial \mathbf{g}} \tr(  \mathbf{W}\mathbf{R}_{yq}^H\diag(\mathbf{g})\mathbf{W}^H)}_{V}  \label{eq:deriv}
    \end{align}

    We have to take the derivative of each term of Eq. (\ref{eq:deriv}). Consider the conversion between matrix notation and index notation and the tricky case of a $\diag(\cdot)$ operator
    \begin{eqnarray}
    [\mathbf{A}\mathbf{B}]_{ik} = \sum_{j} A_{ij}B_{jk}
    \end{eqnarray}
    \begin{eqnarray}
    f=\tr[\mathbf{A}\diag(\mathbf{g})\mathbf{B}] = \sum_{i} \sum_{j} A_{ij}g_{j}B_{ji}
    \end{eqnarray}

    Taking the derivative with respect to the coefficients $g_j$ of the diagonal operator we have

    \begin{eqnarray}
    \frac{\partial f}{\partial g_j} = \sum_{i} A_{ij}B_{ji}=[(\mathbf{A}^T\odot\mathbf{B})\mathbf{1}]_{j}
    \end{eqnarray}

    Therefore, we can write

    \begin{eqnarray}
    \frac{\partial \tr[\mathbf{A}\diag(\mathbf{g})\mathbf{B}]}{\partial \mathbf{g}} = (\mathbf{A}^T\odot\mathbf{B})\mathbf{1}
    \end{eqnarray}

    With these considerations we can take the derivative of terms $I$ , $II$, $III$, $IV$ and $V$ from Eq. (\ref{eq:deriv}). The derivatives of the terms $I$ and $II$ can be computed by

    \begin{eqnarray}
     I =\frac{\partial \tr[\mathbf{R}_{xy}\diag(\mathbf{g})\mathbf{W}^H]}{\partial \mathbf{g}} = [(\mathbf{R}_{xy}^T\odot\mathbf{W}^H)\mathbf{1}] \label{eq:I}
    \end{eqnarray}

    \begin{eqnarray}
    II=\frac{\partial \tr[\mathbf{W}\diag(\mathbf{g})\mathbf{R}_{xy}^H]}{\partial \mathbf{g}} = [(\mathbf{R}_{xy}^H\odot\mathbf{W}^T)\mathbf{1}] \label{eq:II}
    \end{eqnarray}

    To compute the derivative of term $III$ we apply the chain rule

    \begin{eqnarray}
    III=\underbrace{\frac{\partial \tr[\mathbf{W}\diag(\mathbf{g})\mathbf{A}]}{\partial \mathbf{g}}}_{III.1}+\underbrace{\frac{\partial tr[\mathbf{B}\diag(\mathbf{g})\mathbf{W}^H]}{\partial \mathbf{g}}}_{III.2}
    \end{eqnarray}

    \noindent where $\mathbf{A}=\mathbf{R}_{yy}\diag(\mathbf{g})\mathbf{W}^H$ and $\mathbf{B}=\mathbf{W}\diag(\mathbf{g})\mathbf{R}_{yy}$. The term $III.1$ can be computed by
    \begin{eqnarray}
    III.1=[(\mathbf{W}^T\odot(\mathbf{R}_{yy}\diag(\mathbf{g})\mathbf{W}^H))\mathbf{1}] \label{eq:III.1}
    \end{eqnarray}

    \noindent and the term $III.2$ as

    \begin{eqnarray}
    III.2=[((\mathbf{R}_{yy}^T\diag(\mathbf{g})\mathbf{W}^T)\odot\mathbf{W}^H)\mathbf{1}] \label{eq:III.2}
    \end{eqnarray}

    Substituting (\ref{eq:III.1}) and (\ref{eq:III.2}) in (\ref{eq:III}) we have

    \begin{eqnarray}
    III=&&[(\mathbf{W}^T\odot(\mathbf{R}_{yy}\diag(\mathbf{g})\mathbf{W}^H))\mathbf{1}] \nonumber \\
    &+&[((\mathbf{R}_{yy}^T\diag(\mathbf{g})\mathbf{W}^T)\odot\mathbf{W}^H)\mathbf{1}]\label{eq:III}
    \end{eqnarray}

    The derivative of the term $IV$ is given by

    \begin{eqnarray}
    IV=\frac{\partial tr[\mathbf{W}\diag(\mathbf{g})\mathbf{C}]}{\partial \mathbf{g}}=[(\mathbf{W}^T\odot(\mathbf{R}_{yq}\mathbf{W}^H))\mathbf{1}] \textbf{\label{eq:IV}}
    \end{eqnarray}
    \noindent where $\mathbf{C}=\mathbf{R}_{yq}\mathbf{W}^H$. Finally, the derivative of the term $V$ can be computed by

    \begin{eqnarray}
    V=\frac{\partial \tr[\mathbf{D}\diag(\mathbf{g})\mathbf{W}^H]}{\partial \mathbf{g}}= [((\mathbf{R}_{yq}^*\mathbf{W}^T)\odot\mathbf{W}^H)\mathbf{1}] \label{eq:V}
    \end{eqnarray}
    \noindent where $\mathbf{D}=\mathbf{W}\mathbf{R}_{yq}^H$. Substituting (\ref{eq:I}), (\ref{eq:II}), (\ref{eq:III}), (\ref{eq:IV}) and (\ref{eq:V}) in (\ref{eq:deriv}) and equating the derivatives to zero we have
    \begin{align}
    &[\mathbf{W}^T\odot(\mathbf{R}_{yy}\diag(\mathbf{g})\mathbf{W}^H)+(\mathbf{R}_{yy}^T\diag(\mathbf{g})\mathbf{W}^T)\odot\mathbf{W}^H]\mathbf{1}= \nonumber \\
     &\frac{1}{\alpha}([(\mathbf{R}_{xy}^T\odot\mathbf{W}^H)\mathbf{1}]+[(\mathbf{R}_{xy}^H\odot\mathbf{W}^T)\mathbf{1}]+ \nonumber \\
     &-[(\mathbf{W}^T\odot(\mathbf{R}_{yq}\mathbf{W}^H))\mathbf{1}]-[((\mathbf{R}_{yq}^*\mathbf{W}^T)\odot\mathbf{W}^H)\mathbf{1}]) \label{eq:123}
    \end{align}

    To achieve the desired $\mathbf{g}$ we have to do some manipulations with the first term of (\ref{eq:123}). To do this we will write the first and second terms of $\mathbf{g}$ with the index notation and after that we will return to the matrix notation. We can write the first term as
     \begin{eqnarray}
     [(\mathbf{W}^T\odot(\mathbf{R}_{yy}\diag(\mathbf{g})\mathbf{W}^H)\mathbf{1}]=\sum_{j=1}^{KN_T}\sum_{l=1}^{N_R}W_{ji}R_{yy,il}g_{l}W_{lj}^H
     \end{eqnarray}

     \noindent and the second term as
     \begin{eqnarray}
     [(\mathbf{W}^H\odot(\mathbf{R}_{yy}^T\diag(\mathbf{g})\mathbf{W}^T)\mathbf{1}]=\sum_{j=1}^{KN_T}\sum_{l=1}^{N_R}W_{ij}^HR_{yy,li}g_{l}W_{jl}
     \end{eqnarray}

     With some manipulations we can isolate the vector $\mathbf{g}$
     \begin{align}
     &[\mathbf{W}^T\odot(\mathbf{R}_{yy}\diag(\mathbf{g})\mathbf{W}^H)+\mathbf{W}^H\odot(\mathbf{R}_{yy}^T\diag(\mathbf{g})\mathbf{W}^T)]\mathbf{1} \nonumber \\
     &=\sum_{j=1}^{KN_T}\sum_{l=1}^{N_R}W_{ji}R_{yy,il}g_{l}W_{lj}^H+\sum_{j=1}^{KN_T}\sum_{l=1}^{N_R}W_{ij}^HR_{yy,li}g_{l}W_{jl} \nonumber \\
     &=\sum_{l=1}^{N_R} ([(\mathbf{W}^T\mathbf{W}^*) \odot \mathbf{R}_{yy}+(\mathbf{W}^H\mathbf{W}) \odot \mathbf{R}_{yy}^T]_{il})g_{l} \nonumber \\
     &=[(\mathbf{W}^T\mathbf{W}^*) \odot \mathbf{R}_{yy}+(\mathbf{W}^H\mathbf{W}) \odot \mathbf{R}_{yy}^T]\mathbf{g} \label{eq:isolate}
        \end{align}

     Substituting (\ref{eq:isolate}) in (\ref{eq:123}) and solving with respect to $\mathbf{g}$ we have
     \begin{align}
     \mathbf{g}=&[(\mathbf{W}^T\mathbf{W}^*) \odot \mathbf{R}_{yy}+(\mathbf{W}^H\mathbf{W}) \odot \mathbf{R}_{yy}^T]^{-1} \nonumber \\
     &\cdot\frac{2}{\alpha}(Re([(\mathbf{R}_{xy}^T\odot\mathbf{W}^H)\mathbf{1}])- Re([(\mathbf{W}^T\odot\mathbf{R}_{yq}\mathbf{W}^H)\mathbf{1}])) \label{eq:optimumg}
     \end{align}





\section{Clip-level adjustment}

In the following we outline the computation of the clipping factor $\alpha$ based on the signal power. This factor conforms the received signal power between the quantizer range to minimize the overload distortion. The received signal power can be computed by
     \begin{eqnarray}
     P &=& \tr(E[(\mathbf{y}+\mathbf{q})(\mathbf{y}+\mathbf{q})^H]) \nonumber \\
       &=&\tr(\mathbf{R}_{yy}+\mathbf{R}_{yq}+\mathbf{R}_{yq}^H+\mathbf{R}_{qq})
     \end{eqnarray}

\noindent    and received symbol energy by
     \begin{eqnarray}
     E_{rx}=\sqrt{\frac{ \tr(\mathbf{R}_{yy}+\mathbf{R}_{yq}+\mathbf{R}_{yq}^H+\mathbf{R}_{qq})}{N_R}}
     \end{eqnarray}

Thus, the clipping factor $\alpha$ can be obtained from
     \begin{eqnarray}
     \alpha=\beta \cdot\sqrt{\frac{ tr(\mathbf{R}_{yy}+\mathbf{R}_{yq}+\mathbf{R}_{yq}^H+\mathbf{R}_{qq})}{N_R}},
     \end{eqnarray}
\noindent where $\beta$ is a calibration factor. In our simulations the value of $\beta$ was set to $\sqrt{b}$ which corresponds to the quantizer output range, to ensure an optimized performance.

\section{Capacity lower bound}

In \cite{b1} a lower bound on the mutual information between the
input sequence $\mathbf{x}$ and the quantized output sequence
$\mathbf{r}$ of a quantized MIMO system was developed, based on the
MSE approach. We will use a similar procedure to consider a capacity
lower bound of our quantized MU-MIMO system with the optimal AGC and
to derive an expression for computing the achievable rates for the
proposed AGC and LRA-MMSE receiver. We remark that similar analyses
can be considered for the downlink with the use of precoding
techniques \cite{gbd,wlbd,mbthp}. As described in \cite{cover} the
mutual information of this channel can be expressed as
\begin{eqnarray}
I(\mathbf{x},\mathbf{r})=h(\mathbf{x})-h(\mathbf{x}|\mathbf{r})
\end{eqnarray}

Given $\mathbf{R}_{xx}$ under a power contraint $tr(\mathbf{R}_{xx}) \leq P_{\tx}$, we choose $\mathbf{x}$ to be Gaussian, which is not necessarily the capacity achieving distribution for our quantized system. Then, we can obtain a lower bound for $I(\mathbf{x},\mathbf{r})$ (in bit/transmission) as
\begin{eqnarray}
I(\mathbf{x},\mathbf{r})&=&\log_2\det(\mathbf{R}_{xx})-h(\mathbf{x}|\mathbf{r}) \nonumber \\
                       &=&\log_2\det(\mathbf{R}_{xx})-h(\mathbf{x}-\hat{\mathbf{x}}|\mathbf{r}) \nonumber \\
                       &\geq&\log_2\det(\mathbf{R}_{xx})-h(\underbrace{\mathbf{x}-\hat{\mathbf{x}}}_{\epsilon}) \label{eq:23654}\\
                       &\geq&\log_2 \frac{\det(\mathbf{R}_{xx})}{\det(\mathbf{R}_{\epsilon\epsilon})}\label{eq:mutual_information}
\end{eqnarray}

The second term in (\ref{eq:23654}) is upper bounded by the entropy
of a Gaussian random variable whose covariance is equal to the error
covariance matrix $\mathbf{R}_{\epsilon\epsilon}$ of the LRA-MMSE
estimate of $\mathbf{x}$. Thus, we have to compute the expressions
of  $\mathbf{R}_{xx}$ and $\mathbf{R}_{\epsilon\epsilon}$ for our
system. Considering unknown CSI at the transmitter, the
autocorrelation matrix $\mathbf{R}_{xx}$ is given by
\begin{eqnarray}
\mathbf{R}_{xx}=\sigma_x^2\mathbf{I}_{KN_T}, \label{eq:181921}
\end{eqnarray}

\noindent and the error covariance matrix can be computed by
\begin{align}
\mathbf{\mathbf{R}_{\epsilon\epsilon}}=&E[(\mathbf{x}-\hat{\mathbf{x}})(\mathbf{x}-\hat{\mathbf{x}})^H] \nonumber \\
                                     =&\mathbf{R}_{xx}-\mathbf{R}_{xy}\mathbf{G}\mathbf{W}^H-\mathbf{R}_{xq}\mathbf{W}^H-\mathbf{W}\mathbf{G}\mathbf{R}_{xy}^H+ \nonumber \\
                                     &+\mathbf{W}\mathbf{G}\mathbf{R}_{yy}\mathbf{G}\mathbf{W}^H+\mathbf{W}\mathbf{G}\mathbf{R}_{yq}\mathbf{W}^H-\mathbf{W}\mathbf{R}_{xq}^H +\nonumber \\
                                     &+\mathbf{W}\mathbf{R}_{yq}^H\mathbf{G}\mathbf{W}^H+\mathbf{W}\mathbf{R}_{qq}\mathbf{W}^H\label{19245}
\end{align}

Substituting (\ref{eq:181921}) and (\ref{19245}) in
(\ref{eq:mutual_information}) we obtain an expression to compute the
achievable rates for the MU-MIMO system with coarsely quantized
signals.

\section{Results}

To evaluate the results obtained in previous sections we consider a
MU-MIMO system with $K=16$ users who are each equipped with $N_T=2$
transmit antennas and one BS with $N_R=64$ receive antennas. At each
time instant the users transmit data packets with 100 symbols using
BPSK modulation. The channels are obtained with independent and
identically distributed complex Gaussian random variables with zero
mean and unit variance. For each simulation 10000 packets are
transmitted, by each transmit antenna, over a flat-fading channel.
The received signal is quantized with 2, 3, 4 and 5 bits.

\begin{figure}[!h]
\centering
\includegraphics[scale=0.4]{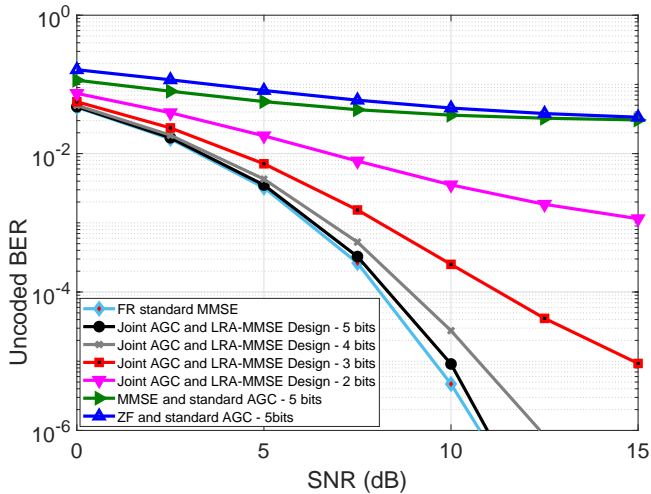}
\caption{Joint AGC and LRA-MMSE receiver performance comparison}
\label{fig:ber}
\end{figure}

Fig. \ref{fig:ber} shows the BER performance of the proposed joint
AGC and LRA-MMSE receiver design. As expected, the standard MMSE
detector achieved, even in a quantized environment, a better
performance than the ZF detector. This occurs because the MMSE
filter incorporates the variance of the receive antenna noise which
improves the accuracy of the MMSE detector at low SNR values.
Moreover, we can see that among all receivers the LRA-MMSE with the
proposed AGC obtained the best performance. The design of this
receiver aggregates the gains by incorporating the AGC and the
effects due to the coarse quantization. The curves also show that,
with the presented approximations, the joint AGC and LRA-MMSE
receiver design achieves a performance very close to the performance
of the Full Resolution standard MMSE receiver (FR standard MMSE).

\begin{figure}[!h]
\centering
\includegraphics[scale=0.425]{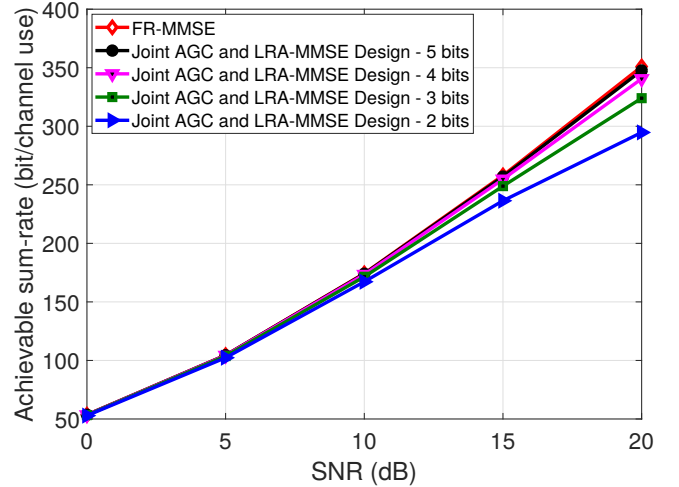}
\caption{Achievable sum-rate of the quantized MU-MIMO system with 32 users with 2 transmit antennas each and one BS with 64 receive antennas.}
\label{fig:capacity}
\end{figure}

In Fig. \ref{fig:capacity} we illustrate the achievable sum rates of
the MU-MIMO system with the joint AGC and LRA-MMSE receiver design
for different numbers of quantization bits. This result shows that,
as the number of quantization bits increases, the sum-rate also
increases approaching those values obtained by the FR standard MMSE
receiver in an unquantized environment.

\section{Conclusions}

In this work we have discussed the joint design of an AGC and a LRA-MMSE receive filter for coarsely quantized MU-MIMO systems. Simulations results have shown that the joint AGC and the LRA-MMSE receiver obtained a performance close to the full resolution MMSE receiver in a quantized MU-MIMO system with 4 and 5 bits of resolution. Furthermore, we have derived an expression for computing the achievable rates for the system. The results have shown that with 4 and 5 quantization bits we achieve a rate very close to the capacity of an unquantized large-scale MU-MIMO system.

\end{document}